\documentclass[12pt,preprint]{aastex}

\usepackage{graphicx}   

\shorttitle{Pulse Width Evolution in X-ray Flares}
\shortauthors{Kocevski et al.}
\def\gtrsim{\mathrel{\hbox{\rlap{\hbox{\lower4pt\hbox{$\sim$}}}\hbox{$>$}}}}
\def\lessim{\mathrel{\hbox{\rlap{\hbox{\lower4pt\hbox{$\sim$}}}\hbox{$<$}}}}

\begin{document}

\title{Pulse Width Evolution of Late Time X-rays Flares in GRBs}

\author{Daniel Kocevski \altaffilmark{1}, Nathaniel Butler \altaffilmark{1,2}, Joshua S. Bloom \altaffilmark{1,3} }

\altaffiltext{1}{Astronomy Department, University of California, 601
Campbell Hall, Berkeley, CA 94720 }
\altaffiltext{2}{Space Sciences Laboratory,
University of California, Berkeley, CA, 94720-7450, USA}
\altaffiltext{3}{Sloan Research Fellow }
\email{kocevski@berkeley.edu, nat@astro.berkeley.edu, jbloom@astro.berkeley.edu}


\begin{abstract}
We study the duration and variability of late time X-ray flares following gamma-ray bursts (GRBs) observed by the narrow field X-ray telescope (XRT) aboard the {\it Swift} spacecraft.  These flares are thought to be indicative of late time activity by the central engine that powers the GRB and produced by means similar to those which produce the prompt emission.  We use a non-parametric procedure to study the overall temporal properties of the flares and a structure function analysis to look for an evolution of the fundamental variability time-scale between the prompt and late time emission.  We find a strong correlation in 28 individual x-ray flares in 18 separate GRBs between the flare duration and their time of peak flux since the GRB trigger.  We also find a qualitative trend of decreasing variability as a function of time since trigger, with a characteristic minimum variability timescale $\Delta t/t=0.1$ for most flares.  The correlation between pulse width and time is consistent with the effects of internal shocks at ever increasing collision radii but could also arise from delayed activity by the central source.  Contemporaneous detections of high energy emission by GLAST could test between these two scenarios, as any late time X-ray emission would undergo inverse Compton scattering as it passes through the external shock.  The profile of this high energy component should depend on the distance between the emitting region and the external shock.

\end{abstract}

\keywords{gamma-rays bursts--- X-rays: general --- high energy:analysis}

\maketitle


\section{Introduction} \label{sec:Introduction}

One of the most unanticipated results to come from the {\it Swift}~spacecraft 
\citep{Gehrels04} is the wide variety of X-ray behaviors observed in the early 
afterglows of gamma-ray bursts (GRBs).
As of January of 2007, {\it Swift}~had detected 206 GRBs and had observed
a subset of $>90$\% of those events with the spacecraft's narrow field
X-ray telescope or XRT \citep{Burrows05a} . Of these events, $>90$\%
show temporal properties that deviate from the simple post cooling break
powerlaw decline that had been seen at late times ($\gtrsim 3 \times
10^{4}$ seconds) by previous spacecraft \citep[e.g.,][]{Frontera00, Gendre06}.
Afterglows with simple powerlaw declines that extend from a few $\sim
10^{2}$ seconds to several days after a burst are seen, for example GRB
061007 \citep{Mundell06}, but they constitute a far minority of the
afterglows observed by the XRT.  Instead, most afterglows show sharp
drops in the observed flux immediately following the gamma-ray emission
\citep{Barthelmy05a}, lasting anywhere from $\sim 10^{2}$ to $\sim 10^{3}$
seconds post trigger.  This is followed by a flattening of the light curve
that can last hundreds of seconds \citep{Granot06} before eventually
transitioning to the late time powerlaw decay previous observed by
other spacecraft.  Most surprisingly, interspersed among these various
components of the prompt afterglow emission have been the detections of
major re-brightening episodes with emission flaring in some cases several
hundred times above the declining afterglow emission \citep{Burrows05b}.
In rare cases, these flares have actually surpassed the luminosity of
the original GRB \citep{Burrows07}.

Numerous papers have been published discussing a variety of mechanisms
that could produce the late time flaring \citep{Zhang06,
Liang06, Falcone06, Mundell06, Perna06, Proga06, Lazzati07, Lee07, Lyutikov06, Fan05}.
Most of these mechanisms place tight constraints on the timescales 
on which the their emission can be produced \citep{Ioka05}.
The simplest explanation would be that the forward shock powering the
afterglow runs into ambient density fluctuations as it moves into
the surrounding medium \citep{Wang00}.  This external shock interpretation
has difficulties explaining the degree of variability that is clearly seen
in many of these flares \citep[e.g.,][and below]{Burrows05b}.  Simple kinematic arguments show that fluctuations due to turbulence of the interstellar medium or variable winds from the progenitor are expected to produce broad and smooth rise and decay profiles, with $\Delta t / t \sim 1$ \citep{Ioka05}. Here $t$ is the time since the gamma-ray trigger and $\Delta t$ is the variability timescale.  Shocks internal to the relativistic outflow \citep{Rees94, Narayan92}, similar to the shocks believed to produced the prompt gamma-ray emission, do not suffer from these same constraints and could in theory produce variability on much shorter timescales.  In the internal shock scenario the rise time of an individual pulse is governed by the time it takes for the reverse shock to propagate back through the shell.  The decay time is largely set by the relativistic kinematics, or curvature effects, in which the arrival of off axis emission from a relativistically expanding shell is delayed and affected by a varying Doppler boost.

Another clue that the flares are produced
in a region distinct from the external shock is that the temporal
decay of the afterglow emission appears largely unaffected by the presence
of flaring.  The temporal index of the afterglow after the flaring activity
is typically consistent with the pre-flare decay index.  Although
most bright flaring occurs within one hour of the GRB, flares have been observed during of each of the light curve phases described above. If, for example, the flare represented the onset of forward or
reverse shock emission of a slow shell catching up and colliding with the
external shock, then these flares would be expected to occur only before
the flat energy injection phase.
Furthermore, \citep{Burrows07} points out flaring in one example of a possible
``naked burst`` \citep{Kumar00}, an event which decays rapidly in
time and therefore exhibits no evidence of external shock emission.  This 
supports the argument that whatever is powering the afterglow is most
likely not creating the X-ray flares, leaving internal shocks or direct
central engine activity as likely methods for their production.

Further evidence that late time X-ray flares might be associated with internal shocks comes from their spectral characteristics.  First, most of the flares are much harder than the underlying afterglow emission and, as reported by \citet{Burrows07}, the spectral characteristics of the afterglow emission appears unaffected by the flaring activity, possibly indicating two distinct emitting regions.  Second, spectral fitting by \citet{Butler07} has shown that many flares can be well fit by the Band model \citep{Band93} that so effectively describes the prompt emission which is largely believed to be the result of internal shock collision.  Furthermore, detailed time resolved spectral fitting of bright flares by \citet{Butler07} has shown that the spectral break energy $E_{pk}$ of the Band model, which represents the energy at which most of the photons are emitted, evolves to lower energy during the flare in a way that is very similar to what is seen in the prompt emission \citep{Norris86}.  The evolution also follows the hardness-intensity correlation \citep{Golenetski83}, a well known relationship observed in the prompt emission that can be attributed to the relativistic effects that produce the decay profile of individual pulses \citep{Kocevski03}.

If the energy released by this activity is converted to radiation
through late time internal shocks, then the question remains as to
the characteristic radius that these internal shocks are occurring as well as the delay in their ejection.
Either the central engine is still functioning and emitting shells at
very late times, or the final few shells of the original outflow, which
were emitted along with the shells that created the prompt emission,
catch up with each other only after a long delay due to a small relative
difference in their bulk Lorentz factor $\Gamma$.  The first scenario
could essentially produce shell collisions at any radius, as the delayed
arrival of the flares would, in this case, primarily reflect the time
that the engine was dormant \citep{Kobayashi97}.  The second scenario, predicts that the late time flares should occur at a radius that is significantly larger than the radius at which the prompt emission
was created, with their delayed arrival being a result of the shells'
time of flight before colliding.  This second scenario leads to a very specific and
testable prediction, namely that the width of individual pulses of
emission should become broader and less variable when originating from
shells of increasing collision radii $R_{c}$.  \citet{Ramirez-Ruiz00}
tested for this pulse width evolution in the light curve
profiles of BATSE events and found no evidence for any such effect.
They concluded that the  prompt emission observed by BATSE must have
been produced over a small range of $R_{c}$ from the central engine and
that no significant deceleration of $\Gamma$ could have occurred over
the duration of the observed activity.

The goal of this paper is to extend the gamma-ray pulse width analysis to the late time flaring X-ray emission following GRBs. The public catalog of {\it Swift} XRT flares (see also Chincarini et al. 2007) represents the first dataset to test the internal vs. external shock scenario for this flaring activity.
Whereas previous studies were limited to prompt emission occurring less than 100 seconds after trigger, the late time X-ray flares give us the opportunity to test for pulse width and variability evolution out to, in some case, 1000 seconds after the trigger of the GRB where this effect may be more pronounced.  We provide a simple derivation of the expected pulse width evolution in both small $\Delta\Gamma$ and delayed engine activity scenarios in $\S 2$, followed by a discussion of our data reduction techniques in $\S 3$ and results in $\S 4$.  We find evidence for pulse width evolution in 28 flares as well as a qualitative trend of decreasing variability as a function of the flare's time of peak flux.  We discuss the implications of our observations in $\S 5$.  This work expands upon and formalizes our previous reports \citep{2006AAS20922703K,ButlerGLAST07} of the discovery of pulse width evolution.

\section{Pulse Width Evolution} \label{sec:PulseWidthEvolution}

The standard fireball model postulates the release of a large amount of
energy by a central engine into a concentrated volume \citep{Cavallo78},
which causes the resulting outflow to expand and quickly become
relativistic \citep{Paczynski86}.  In the internal shock scenario
\citep{Rees94}, this outflow is assumed to be variable, consisting of
multiple shells of differing bulk Lorentz factors $\Gamma$.  These shells
propagate and expand adiabatically until a faster shell collides with
a slower one, causing the shells to coalesce and convert a significant
fraction of their kinetic energy into radiation, most probably through
optically thin synchrotron radiation.  The resulting pulse profile that
is observed is a convolution of two distinct timescales.
The rise time of the pulse is largely due to the time it takes for
the reverse shock that is induced by the collision to cross the width
of the faster shell.  The decay time, on the other hand, is governed
mainly by angular and kinematic effects where off axis emission is
delayed and effected by a varying Doppler boost due to the curvature
of the relativistic shell (see Figure 1 in \citet{Kocevski03}).  As a
result, the decay time can be, and in most cases is, much longer than
the rise time, leading to an asymmetric pulse profile.  The combination
of these two timescales (the shell crossing time and the angular time)
naturally explains the so called ``fast rise exponential decay'' or FRED
pulses that are so ubiquitous in prompt GRB emission.\footnote{Here we
assume that the intrinsic cooling time $\Delta t_{c}$ of the shell is
insignificant compared to the duration of the shell crossing  $\Delta
t_{r}$ and angular  $\Delta t_{d}$ timescales because of the magnetic
field strength required to produce the gamma-ray emission}

If we examine these two timescales in more detail, we can see that
the rise time is primarily a thickness effect and can be expressed as
$\Delta t_{rise} = \delta R / c(\beta_{2}-\beta_{rs})$, where $\delta R$
and $\beta_{2}$ are the thickness and velocity of the second shell that
is catching up to the first and $\beta_{rs}$ is the velocity of the
reverse shock.  If both the slow and fast shells have Lorentz factors
of roughly the same order $\sim \Gamma$, then the resulting rise time is
of order $\sim \delta R / c$.  Because the merging shells are traveling
forward at a velocity very close to the speed of light ($\Gamma \gg
1$), the resulting coalesced shell keeps up with the photons that
it emits.  Therefore, any emission activity over a fixed duration will
appear to an outside observer to be compressed in time by a factor of
$1/2\Gamma_{m}^{2}$, where $\Gamma_{m}$ is the resulting Lorentz factor
of the merged shell.  The observed rise time can therefore be written as
\begin{equation} \label{eq:rise}
	\Delta t_{r} \approx \frac{\delta R }{2c\Gamma_{m}^{2} }
\end{equation}
So given a sufficiently large $\Gamma_{m}$, internal shocks can
essentially produce variability along the line of sight on arbitrarily
short timescales.

Angular (or curvature) effects have the opposite effect, causing a
broadening of the overall emission profile that can quickly come to
dominate the observed pulse shape.  The decay timescale is essentially
the difference in light-travel time between photons emitted along the
line of sight and photons
emitted at an angle $\theta$ along a shell of radius $R$.  This can be
stated as
\begin{equation} \label{eq:decay}
	\Delta t_{d} = \frac{R(1-\cos\Delta\theta)}{c} \approx
	\frac{R(\Delta\theta)^{2}}{2c} \approx \frac{R}{2c\Gamma^{2}}
\end{equation}
Where the last step assumes that the shell is moving with sufficient
velocity such that the solid angle accessible to the observer is
limited by relativistic beaming and thus given by $\Delta\theta \sim
1/\Gamma^{2}$. Therefore comparing Equation \ref{eq:rise} and Equation
\ref{eq:decay}, we can see that curvature effects become important
whenever the radius of the shell exceeds the shell thickness, which is
true for all but the earliest moments of the shell's expansion.

The significance of Equation \ref{eq:decay} is that angular effects should
scale linearly with the radius of the emitting shell and therefore pulse
durations should become broader as shell collisions occur further from
the central engine.  If the flares are the result of multiple shells that have been ejected almost instantaneously (or at least within a timescale that is small compared to the overall GRB
duration) but collide at very late times due to a small dispersion
in Lorentz factors, then one would expect that these late collisions would occur at greater radii.
In this scenario, we can replace the radius of the shell in Equation
\ref{eq:decay} with the time $t$ since the ejection of the first shell
by noting that the observed radius of a spherical shell expanding with $v
\sim c$ can be approximated as $R \approx ct\Gamma^{2}$, where the extra
factor of $\Gamma^{2}$ is due to relativistic corrections, leading to
\begin{equation} \label{eq:decay_t}
	\Delta t_{d} \approx \frac{t}{2}
\end{equation}
Therefore, the late shock scenario would predict a linear correlation
between a shell's time of flight and the resulting pulse duration,
independent of the Lorentz factor of the shell.  This relationship
between the pulse duration the time since the ejection of the internal
shocks has been noted before. \citet{Fenimore96} found, through a much
more detailed derivation, that a pulse's {\em FWHM} should scale roughly
as $0.26T_{0}$ to  $0.19T_{0}$ as the low energy powerlaw index $\alpha$
varies from 1 to 2.  Similarly, \citet{Ioka05} derive that the variability
of flares that result from refreshed shocks should be limited by $\Delta
t \geqslant t_{p}/4$.  In each case, flares occurring at larger radii
are expected to produce broader pulse durations.

This relationship between $\Delta t_{d}$ and $t$ is modified if there is
an intrinsic delay $\Delta t_{engine}$ in the ejection of the subsequent
shells by the central engine.  If we imagine two shells emitting at time
zero and time $\Delta t_{\rm engine}$, provided the Lorentz factor of
the second shell $\Gamma_2 > \Gamma_1$, the Lorentz factor of the first
shell, the shells will collide at time

\begin{equation}
t_c = { \Gamma_1^2 \Delta t_{\rm engine} \over \Gamma_2^2-\Gamma_1^2}
\end{equation}

\noindent If the the shells have equal mass, which corresponds to the
maximal efficiency for conversion of kinetic energy into radiation,
energy and momentum conservation lead to a merged shell with Lorentz
factor $\Gamma_m = \sqrt{\Gamma_1 \Gamma_2}$.  The timescale over which
the shell emits will be governed by the longest timescale of $\Delta
t_a$, $\Delta t_r$, or $\Delta t_c$, the angular, the radial, or the cooling timescale, respectively.  The angular
and radial timescales can both be given as:

\begin{equation}
\Delta t_a \approx \Delta t_r \approx { R \over 2c \Gamma_m^2 } = {
R \over 2c \Gamma_1 \Gamma_2}.
\end{equation}

\noindent The time at which a flare is observed will be $t_c$, and the
observed duration will be $\Delta t = t_c \Gamma_1/\Gamma_2 \approx
t_c/2$ for an efficient collision with $\Gamma_2=2\Gamma_1$.  For this
$\Gamma_2/\Gamma_1$ ratio, the flare duration $\Delta t$ is related to
the duration at the central engine by $\Delta t = \Delta t_{\rm engine}
/6 \sim \Delta t_{\rm engine}$. Therefore, if there is any appreciable
delay in the ejection of relativistic material from the central engine,
the resulting pulse shape will not necessarily reflect the shell radius,
but rather the intrinsic delay between the ejection of the two shells.
Any correlation between pulse shape and time of peak flux must then be
attributed to the activity of the central engine.

\section{Data $\&$ Analysis} \label{sec:Data}

We select a subsample of 28 bright ($\gtrsim 10$ cts/s) flares that are
fully time-sampled (i.e., no gaps in their light curves) in 18 separate
GRB afterglows observed by XRT. The Burst Alert Telescope (BAT)
and XRT data were downloaded from the {\it
Swift}~Archive\footnote{ftp://legacy.gsfc.nasa.gov/swift/data}
and processed with version 0.10.3 of
the {\tt xrtpipeline} reduction script and other tools from the
HEAsoft~6.0.6\footnote{http://heasarc.gsfc.nasa.gov/docs/software/lheasoft/}
software release.  We employ the latest (2006-12-19) calibration
files available to us at the time of writing.

The reduction from cleaned event lists output by the {\tt xrtpipeline}
code and from the HEAsoft BAT software
to science ready light curves and spectra is described in detail
in \citet{Butler07}.  The bright XRT flare data are taken overwhelmingly
in windowed-timing (WT) mode, which mandates special attention to bad
detector columns.  As the spacecraft moves, a significant and time
varying fraction of the source flux can be lost if source counts fall
on the bad columns.   To account for this \citep[see,][]{Butler07},
we calculate exposure maps for the WT mode on a frame-by-frame
basis. We accumulate 0.3-10.0 keV counts in each light curve bin
until a fixed signal-to-noise ($S/N$) of 3 is achieved. All of the
resulting light curves to which we apply our analysis are publicly
available\footnote{http://astro.berkeley.edu/$\sim$nat/swift}. A composite
light curve plot showing all 18 GRBs in our data set is shown in Figure 1.

\subsection{Flare Duration Measures}

The first step in our analysis consists of measuring the global flare
duration timescales.  Because we have found no one functional form (e.g.,
Gaussians) to adequately fit the X-ray flare time profiles (Figure 1),
we employ non-parametric duration estimators. We consider the flare
$T_{90}$ duration as the time required to accumulate between 5\% and 95\%
of the flare counts.  We also define a rise time as the time between 5\%
accumulation and the time of count rate peak. Errors on these quantities
are determined from the non-parametric bootstrap (i.e., by recalculating
the quantities for data simulated using the measured data and errors).

A bias affecting these duration measures (and probably all durations
measures) is the unknown background under the flares.  As discussed above,
studies have shown that a powerlaw decaying background likely does exist.
However, it cannot cleanly be measured in many of our events and not at
all for events which suffer from data gaps.  In an effort to avoid such
biases in our duration measurements, we have restricted our analysis to
flares that are typically 2$-$3 orders of magnitude above background.

\subsection{Flare Variability Measures}

The flare duration and the component rise and decay times are gross
measures of variability.  In addition to this information, we attempt
to measure the finer timescale fluctuations in the light curves which
may prove important for inferring the size and nature of the flare's
emitting region.  Several methods for measuring signal power versus time
scale have been applied to astronomical inquires and in GRB research in
particular. Several authors (e.g., Belli 1992; Giblin, Kouveliotou, \& van
Paradijs 1998; Beloborodov, Stern, \& Svensson 2000) employ the Fourier
power spectral density (PSD) to study time variations in GRB light curves.
The autocorrelation function (ACF), which is simply the Fourier transform
of this PSD, has been used to demonstrate a narrowing of GRB pulses with
increasing energy band (e.g., Fenimore et al. 1995). Below, we utilize
the {\em first order structure function}, which is directly proportional to
the ACF and has a rich heritage in the study of quasar time histories
(e. g., Simonetti, et al. 1985; Hughes, Aller, \& Aller 1992).

Because the light curves of flaring sources are by definition
non-stationary signals (i.e. signals whose frequency content changes
with time) which exhibit sharp discontinuities, Fourier transforms do a
particularly poor job of accurately measuring their power on both short
and long timescales.  Furthermore, they offer no ability to distinguish
the temporal variations of specific spectral components (i.e. the time at which a characteristic frequency changes in a light curve)  They are also somewhat more prone than $ACF$ methods to aliasing effects due to irregularly time-sampled data.

Instead of constructing a PSD using superpositions of sines and cosines,
we can perform the equivalent analysis by constructing a scaleogram
through the use of a discrete Haar wavelet transform.  The Haar wavelet
is the simplest possible wavelet, consisting of a step function, and has
been previously exploited to ``denoise'' GRB light curves (e.g., Kolaczyk
\& Dixon 2000) and to infer milli-second variability during the first
seconds of bright BATSE GRB \citep{Walker00}.

As described in more detail in the Appendix, we calculate the structure
function from Haar wavelet coefficients as:
\begin{equation}
\sigma^2_{X,\Delta t} = \Delta t/t \sum_{i=0}^{t/2\Delta t-1} 
(\bar X_{2i+1,\Delta t} - \bar X_{2i,\Delta t})^2.
\end{equation}
where $X_i$ is the natural logarithm of the observed XRT count
rate in bin $i$ at time $t$, and $\Delta t$ is the timescale (or time
``lag'') between successive bins.  The bar over the $X_i$ denotes an
averaging with respect to shorter timescales, which is accomplished by
the discrete wavelet transform (see, e.g., Press et al. 1992).	If this
averaging were not performed, $\sigma^2_{X,\Delta t}$ would be equal to
the structure function $SF = < (X_{i+\Delta t}-X_i)^2 > =
1 - ACF$.  Instead, we have an estimator for $SF$, which ends up being
far easier to interpret, as we discuss in the Appendix.

\section{Results} \label{sec:Results}

\subsection{Pulse Broadening in an Individual Event}

A composite BAT and XRT light curve for GRB 060714 is shown in Figure 2.
The red solid line represents a multiply-broken powerlaw fit to the
light curve.  The inflection points in the fit allow us to measure the
boundaries and durations of the individual pulses within the signal. Each
pulse is delineated by the short-dotted lines with the corresponding
pulse duration labeled below the light curve.  A general trend can be
seen in which the pulse durations become broader as the burst progresses,
with the shortest activity occurring early in the event.

The bottom panel plots the minimum timescale $\Delta t$ for which
$\sigma_{X,\Delta t}$ is at least $3\sigma$ above the floor expected
from Poissonian fluctuations.  Consistent with the trend seen in pulse
duration, this minimum variability timescale, which is calculated without
fitting the data, increases roughly as a powerlaw as the burst moves
from early gamma-ray emission to late X-ray emission. To show that this
increase in the variability timescale is not simply due to an increase in
the data binning as the burst fades, we have plotted the time binning as a
short dotted line in the bottom panel.  For most of the event ($t\lesssim$
200 seconds), the timescale below which little or no significant power
exists is at least an order of magnitude higher than the resolution
of the light curve allowed by the binning of the signal. Overall, the
bottom panel shows that the very fast time variability associated with
the prompt GRB emission dies out at late times.

Although the general broadening of the pulse durations seen in Figure
2 can be detected within the separate BAT and XRT light curves,
the comparison of the pulse durations can only be qualitative when
considering a light curve that spans both detectors.  This is because
GRBs are typically wider at lower energies \citep{Fenimore96}, a direct
result of the evolution of their spectral break energy $E_{pk}$ to lower
energies.  We shown in \citet{Butler07} that the X-ray flares typically 
have $E_{pk}$ in the X-ray band, while the earlier GRB emission has $E_{pk}$ in the
gamma-ray band. Therefore pulses are expected to be intrinsically broader
in the 0.3$-$10.0 keV bandpass of the XRT than the higher 10$-$100 keV
observed by BAT.

\subsection{Pulse Broadening in the Sample Taken as a Whole}

To eliminate the pulse broadening between separate energy bands, we
limit our quantitative comparison of pulse durations (both within a
single GRB and across our entire sample) to measurements made using
only the XRT data on each event. This  comparison is shown in Figure 3,
where we plot pulse duration versus time of peak flux for our entire
sample of XRT observed GRBs.  The flares associated with each GRB are
represented by the same color and symbol, with several GRBs exhibiting
multiple flares throughout their early afterglow.

As a whole, the sample shows a clear correlation between the flare duration and the time of peak flux since the GRB trigger.  The resulting correlation strength is Kendall's $\tau_{K} = 0.7$, with a significance of $10^{-7}$. The slope is consistent with linear, implying $\Delta t \propto t_{p}$, which  cancels out the effects of cosmological redshift.  We find no significant correlation between duration and redshift ($\tau_K=0.2$, signif.$=0.2$), further ruling out cosmological time dilation as the source of this correlation.

Roughly, half of the events with multiple flares (those plotted in color
in Figure 3) show a trend toward increasing duration with observation
time.  The other half show an anticorrelation.	The pulse durations,
time of peak flux, and rise times for al the flares in our sample can
be found in Table 1.

\subsection{Haar Structure Function View}

Figure 4 shows $\sigma_{X,\Delta t}$ versus $\Delta t$ and
$\sigma_{X,\Delta t}$ versus $\Delta t/t$ for the ensemble of flares under
study\footnote{We reserve a more detailed study of the Haar structure
functions of individual flare events and {\it Swift} GRBs in a separate
paper \citep{Butler07a}.}.  In this scaleogram plot,
we show only $3\sigma$ excesses over the power associated with Poisson
fluctuations and report lower values as $3\sigma$ upper limits.

An X-ray flare is an emission episode uncorrelated in time to the
afterglow flux prior to and after the flare. During the flare and on
timescales short relative to the flare duration, the flux will be highly
correlated in time and there will be a linear rise in $\sigma_{X,\Delta
t}$.  This can be observed to arbitrarily short timescale if the fading
powerlaw tail of a flare is measured with very high $S/N$.  On the other
hand, as we describe in more detail in the appendix, correlated behavior
in the light curve flattens the structure function, and this provides 
a direct measure of the flare timescales.

Consistent with the pulse duration correlation seen in Figure 3, the scaleogram plots show a range of important flaring timescale $dt=30-300$s, which becomes much tighter in units $dt/t=0.1-0.5$.  We observe a minimum characteristic timescale $dt/t=0.1$. 

The fractional flux variation levels at the minimum timescale are large ($\sigma_{X,\Delta t} \gtrsim 80$\%), suggesting that the variations correspond to gross features in the light curve.  Consistent with this interpretation, we observe the flare rise times to have $\Delta t{\rm rise}/t=0.1$ on average (Figure \ref{fig:time_plot_rise}), and it is likely the sharp flare rises which produce the shortest timescales reflected in the structure function turnover.  From the linear $\sigma_{X,\Delta t}$, we can rule out significant flickering on timescales shorter than $dt/t=0.1$ (or $dt=30$s) at very small $\gtrsim 3$\% fractional flux levels (Figure 3).  We discuss the flare noise properties as a function of timescale in more detail in \citet{Butler07a}.

For observation times in the 100 to 1000 second range, $dt/t=0.1$ implies emission radii $R_c\approx 10^{15}$ cm$-$$10^{16}$ cm, for a bulk Lorentz factor $\Gamma=100$ (Equation \ref{eq:rise}).  The observable emission is restricted to an angle $\approx 1/\Gamma$, implying an effective emitting region of size $\delta R\approx R_c/\Gamma \approx 10^{14}$ cm$-$$10^{15}$ cm, compared to the typical external shock values of $10^{16}$ cm in the first hour or so \citep{Piran99}.

\section{Discussion} \label{sec:Discussion}

The results from the temporal analysis outlined above provide substantial
evidence that both the pulse duration and pulse variability of late time
X-ray flares evolve with time. Both the pulse duration and variability
timescales appear to have a narrow intrinsic range in $\Delta t/t =
0.3\pm 0.2$, consistent with a narrow range found independently by
\citet{Burrows07} for $\Delta t_{\rm rise}$ and by \citet{Chincarini07} for
$T_{90}$.  GRB~060714 provides the best example of this
behavior in an individual event.  Several other individual GRBs display a similar increasing pulse duration trend
among their associated flares, although several bursts
do not (e.g., GRB 060210).  For the bursts with multiple flares, only half
show increasing flare durations.  Each burst event typically shows only 1,
sometimes 2 (and 3 in one case) separate flares.  These multiple flares
within individual GRBs are only weakly separated in logarithmic time,
and hence probe a small range of $R_{c}$ or  $t_{engine}$ which may not allow for a clean measurement of time evolution in individual events.

The linear relationship between $\Delta t$ and $t_{p}$ is consistent with the pulse width evolution that is expected from the angular effects of late internal shocks at large radii as outlined in $\S$\ref{sec:PulseWidthEvolution}.  Because we do not see a significant alteration of the afterglow light curve after the occurrence of an X-ray flare, the standard refreshed shock model, in which the trailing shells catch up to the leading shell only after the leading shell decelerates due to an external medium, is disfavored.  Although such a scenario is expected to produce a correlation between the pulse width and time of peak flux on the order of $\Delta t \geqslant t_{p}/4$ \citep{Ioka05}, the trailing shells should have the effect of increasing the overall afterglow energy and thus have a discernible effect on the afterglow light curve, which is not seen. Therefore, the internal shocks producing the flares would have to be occurring behind the leading shock that has begun powering the afterglow, with their late occurrence, in this scenario, being due to a small relative Lorentz factor between the two inner shells. 

The primary difficulty with this interpretation is the high flux ratio between the prompt and late emission, given the relatively small $\Delta\Gamma$ needed to explain the late collision time.  As shown in detail by \citet{Krimm07}, the efficiency $\epsilon$ of an internal shock in converting a system's kinetic energy into radiation scales roughly as  $\epsilon \sim \Delta\Gamma^{2}$, so the observed flux drops quickly as the contrast between the Lorentz factor of the shocks decreases.  This posses a problem for the flares observed by Swift, as many exhibit peak fluxes that are significant fractions of, and in some cases comparable to, their associated prompt emission.  The small $\Delta\Gamma$ scenario would require an extremely large total amount of kinetic energy to remain in the system after the release of the prompt emission, given the low efficiency of the late collisions.  These late and highly energetic shocks would, after producing the flaring activity, eventually collide with the external shocks and affect the observed afterglow light curve, something that is not seen in all events with flares. 

Alternatively, if the late nature of the X-ray flares is due to a significant delay in the ejection of late shells by the central engine, then the necessity of a small $\Delta \Gamma$ is eliminated, alleviating this efficiency constraint.  As described in $\S$\ref{sec:PulseWidthEvolution}, the arrival time $t_{c}$ and pulse width $\Delta t$ would then directly reflect the activity of the central engine.  Therefore, in this scenario, the correlation between $t_{c}$ and $\Delta t$ would require an explanation intrinsic to the powering and/or reactivation of the central engine at late times.  Several authors have suggested mechanisms by which the central engine could be active at late times, most involving late-time fallback material or a long lived accretion disks around a central black hole.  A model proposed by \citet{King05} suggests that the late-time activity could be attributed to the fragmentation and accretion of a collapsed stellar core resulting in a sporadic release of energy rather than the classic view of a single cataclysmic event.  Similarly, \citet{Perna06} have proposed a viscous disk model in which the late-time activity is due to re-energization by material that falls in from a range of initial radii toward the accreting black hole.  In this scenario, the correlation between $t_{c}$ and $\Delta t$ would be due to the range of radii from which the accreting material was falling.  Material at large radii, if continuously distributed throughout its orbit, would take longer to fall back onto the central black hole and would do so over a longer duration, due to its larger orbital circumference.  

These models are not without their own share of difficulties.  The simple fragmentation models \citep{King05} are inconsistent with the implication of the spectral evolution seen in may flares \citep{Krimm07}. 
Similarly, the viscous disk model requires a continuous distribution of material at discrete orbits to account for the episodic nature of the flares as well as an extremely long lived, and hence low viscosity, accretion disk to explain flares at 1000 seconds after the original collapse.  

It cannot be completely ruled out that the observed time evolution is due
to spectral evolution or the superposition of multiple flares. Consider
060124 \citep{Butler07, Romano06}, in which the flares may in fact be
the prompt emission, because the faint BAT trigger may be a pre-cursor.
At high energies, the first XRT ``flare'' resolves into 2$-$3 shorter
timescale BAT flares, which are blurred together in the XRT.  We note
that a shift of time origin for 060124 from $t\sim 0$ s to $t\sim 300$ s,
corresponding to a shift in origin from the pre-cursor to the flare start,
does not lead to a violation of the $\Delta t$ and $t_{p}$ correlation.
Although, if we used the BAT flare durations, the correlation could
be violated.

This indicates that spectral considerations are important, and that we are
likely measuring in the XRT (in some cases) a pulse superposition. The
duration which increases in time still appears to measure the duration
of major emission activity, however, it is not clear that these are
individual pulses.  We know that spectra of late time flares are evolving
strongly \citep{Butler07} during the flares.  However, we observe only
a weak correlation between peak time and hardness, indicating that there
is a diversity of flare spectra at each epoch.

Another important concern involves the powerlaw background onto which most of these flares are superimposed.	 Although we have not attempted to subtract the background from the events in our sample (because the backgrounds are not well defined), this should not dominate the observed correlation.  We have selected the brightest flares for analysis, which have peak fluxes orders of magnitude greater than the underlying background flux.  The correlation is also strong for measures of duration like $T_{50}$ or the \citet{Reichart01} $T_{45}$, which are largely insensitive to pulse tails.  Finally, we note that the flare rise time also strongly correlates with the peak time $t_{p}$ as shown in Figure 5.

Barring any of these selection and/or analysis effects and assuming that the pulse width evolution is real, one possible test to distinguish between the late internal shocks with small contrasts $\Delta\Gamma$ and direct central engine activity may come from contemporaneous high energy emission during the X-ray flares.  If the internal shocks creating the flares are occurring behind the external shock, then one would expect the X-ray photons to be boosted to higher energies by a factor of $\Gamma_{FS}^{2}$ through inverse Compton (IC) scattering as they pass through the external shock \citep{Rybicki79}.  The soft X-ray 10 KeV photons associated with the X-ray flares could easily be boosted into the 1$-$100 MeV range depending on the Lorentz factor of the external shock.  The temporal profile of this high energy component should depend heavily on the distance behind the external shock at which this emission originated \citep{Wang06}, as the duration of the IC component will reflect the geometry of the external shock, roughly $R/2\Gamma^{2}c$.  The ratio between the flare duration and the IC component's duration should approach 1:1 as the radius of the internal shock producing the flare approaches the external shock radius.  Internal shocks that result from delayed central engine activity do not necessarily have to be at large radii to produce the longer observed durations.  Therefore, larger IC component to flare duration ratios are expected for flares produced from small radii collisions.  Even if these late-time collisions at small radii have intrinsically longer durations, as suggested by late central engine activity models, the additional light travel time from the origin of the late time flares to the external shock as it expands may make this change in duration ratios measurable.  Such a test for contemporaneous high energy emission will be aptly suited for the upcoming GLAST mission which will be sensitive to photons up to $>$ 300 GeV.

\section{Acknowledgments} \label{sec:acknowledgments}

D.K. acknowledges financial supported through the NSF Astronomy $\&$ Astrophysics Postdoctoral Fellowships under award AST-0502502. N.B. gratefully acknowledges support from a Townes Fellowship at U.~C. Berkeley Space Sciences Laboratory and partial support from J. Bloom and A. Filippenko.  J. S. B. and his group are partially supported by a DOE SciDAC Program through the collaborative agreement DE-FC02-06ER41438.  We also thank Phil Chang, Edison Liang, and Demos Kazanas for their thoughtful discussion. 


\section{Appendix}

We describe here the mathematical representation of a Haar wavelet and
its use in the construction of a scaleogram closely related the the $ACF$
and first order structure function $SF$.

Given $T$
successive data bins $X_i$, we define the Haar wavelet coefficients
$h_{i,1}$ on scale $\Delta t=1$ as
\begin{equation} \label{eq:hi1}
h_{i,1} = X_{2i+1}-X_{2i},  \quad i=0,...,T/2-1.
\end{equation}
At the same time, we can calculate the signal smoothed over
a 2 bin scale $\Delta t=2$:
\begin{equation}
\bar X_{i,2} = {1 \over 2}(X_{2i+1}+X_{2i}),  \quad i=0,...,T/2-1.
\end{equation}
By successively differencing and smoothing the signal on dyadic
scales $\Delta t=1,2,4,$ etc., we build up the discrete Haar transform
(see, also, Press et al. 1992):
\begin{equation} \label{eq:hidt}
h_{i,\Delta t} = \bar X_{2i+1,\Delta t}- \bar X_{2i,\Delta t},	\quad
i=0,...,T/2\Delta t-1.
\end{equation}
If the $X_i$ are uncorrelated with equal variance, then the
$h_{i,\Delta t}$ will be approximately linearly independent. We form a
Haar scaleogram by averaging the $h_{i,\Delta t}$ at each scale
$\Delta t$:
\begin{equation}
\sigma^2_{X,\Delta t} = \Delta t/t \sum_{i=0}^{t/2\Delta t-1} h_{i,\Delta
t}^2 = \Delta t/t \sum_{i=0}^{t/2\Delta t-1} (\bar X_{2i+1,\Delta t} -
\bar X_{2i,\Delta t})^2.
\end{equation}
In practice, we calculate this average as an average weighted by the
data measurement uncertainties, $w_i=1/\sigma_{D,i}^2$.

This quantity, also known as the Allan (1966) variance, is closely
related to the structure function $SF = < (X_{i+\Delta
t}-X_i)^2 >$, where $<...>$ denotes an average over the data.  Unlike
$\sigma^2_{X,\Delta t}$ the quantity $SF$ is calculated without averaging
the data on scale $\Delta t$ before differencing on that scale. This
leads to a scaleogram with correlations (even for uncorrelated input data)
between nearby data bins.  The uncorrelated scaleogram $\sigma^2_{X,\Delta
t}$ is therefore easier to fit and interpret, while both scaleograms
have similar shapes for a wide variety of noise models.

\subsection{Flare Ensemble Haar Structure Function}

Because the Haar wavelets encode signal scale information as a function of
time, it is possible to calculate $\sigma^2_{X,\Delta t}$ for arbitrary
time sections of a light curve (e.g., Figure 2) or for the full light curve.

To make useful scaleogram plots for multiple GRB flares (e.g., Figures 4),
we place the times series data end-to-end and perform the
Haar transform as though the data were binned on an even time grid.
Transform coefficients formed by differencing data from separate events
are discarded. By saving the actual time since GRB trigger $t$ and time
bin width $\Delta t$ for each wavelet coefficient, we can then rebin
the coefficients in time on a dyadic grid starting with the minimum bin
size.  In this fashion, it is possible to plot statistically independent
$\sigma_{X,\Delta t}$ points versus the physically meaningful $\Delta t$
or $\Delta t/t$.

For $X_i$ in Equations \ref{eq:hi1}$-$\ref{eq:hidt}, we use the natural
logarithm of the XRT count rate.  Because the counts have been binned
to a fixed $S/N$ ratio, the error in $X_i$ is approximately constant
($\sigma_D \approx 1/3$). The natural logarithm is also useful because
powerlaw flux variations lead to a ``zero-flaring'' scaleogram with
$\sigma_{X,\Delta t} \propto \Delta t$, as can be seen from a Taylor
expansion of the flux in time. Also because we are working with the
logarithm of the count rate, $\sigma_{X,\Delta t}$ can be interpreted
as a root-mean-square (RMS) fractional variation in the flux $F$ (i.e.,
$\delta X \approx \delta F/F$).

\subsection{Structure Function Interpretation}

Following the discussion in Hughes, Aller, \& Aller (1992), 
on short timescales, the scaleogram $\sigma_{X,\Delta t}$ asymptotes
to $\sigma_D$, where $\sigma_D$ is the data measurement uncertainty.
Because we know $\sigma_D$, we can subtract this flattening out. (This is
typically not possible for $SF$ due to the introduction of correlations
in the data.) From the Cauchy-Schwarz inequality \footnote{Recall that
the Cauchy-Schwarz inequality states that $| \langle x,y \rangle
|^{2} \leqslant \langle x,x \rangle \cdot \langle y,y \rangle$ and that
the two sides are equal only if x and y are linearly dependent.}, the
scaleogram increases with increasing time lag. It eventually saturates
to a characteristic signal level $\sigma_{\rm signal}$ at time $\lessim
T_{90}$, once we begin to run out of correlated  variations in the signal.

On intermediate timescales, the slope of $\sigma_{X,\Delta t}$ depends
on the shape of the light curve and on the noise spectrum of possible
low-level or unresolved flares. If the light curve is correlated
on these timescales, which is to say smooth on these timescales,
$\sigma_{X,\Delta t}$ will increase as $\Delta t$. If, however, the
light curve is dominated by the sum of slowly decaying responses to low
level flares, a characteristic ``flicker noise'' spectrum ($PSD(f)\propto
1/f$) may result and $\sigma_{X,\Delta t} \propto \Delta t^0$.	Hence,
we can test for flaring as a function of timescale by measuring powerlaw
$\sigma_{X,\Delta t}$ slopes less than unity.

The fading powerlaw tail of a flare measured with infinite $S/N$ would
produce a statistically significant $\sigma_{X,\Delta t}$ for arbitrarily
small $\Delta t$. These timescales, where $\sigma_{X,\Delta t} \propto
\Delta t$, are therefore uninteresting.  However, the beginning of
a $\sigma_{X,\Delta t} \propto \Delta t^0$ phase yields a physically
meaningful timescale for the flaring. The breadth of this phase indicates
the range of $\Delta t$ present in the light curve.



\section{Figure Captions}

{\bf Fig. 1.} - The XRT count rate (cts/s) plotted vs. time since trigger
for all 28 flares in 18 separate GRBs. The light curves in this plot
are rebinned to $S/N=10$.  A qualitative trend between pulse width and time of peak flux can be seen by inspection.

{\bf Fig. 2.} - {\it Top Panel.} A composite BAT and XRT light curve
for GRB 060714 showing an increasing pulse duration as a function of
time. {\it Bottom Panel.} The minimum variability timescale in the
composite light curve (with power that is at least 3$\sigma$ above
that which is expected from Poissonian fluctuations).  The variability of
the light curve increases with time, roughly as a powerlaw of $\Delta
T_{min} \propto T^{1.9 \pm 0.6}$.

{\bf Fig. 3.} - The pulse duration $T_{90}$ versus time of peak flux
$T_{p}$ for our entire sample of XRT observed flares.  Multiple flares from
individual GRBs are displayed with a unique color-symbol combination, whereas GRBs with only one flare are represented by a black diamond.  A strong trend ($tau_{K} = 0.7$) between
pulse width and the time since trigger, as measured in the observer
frame, is clear from the data. Only half of the GRBs with multiple flares
display a similar increasing pulse duration trend between their associated
flares.  We conclude that the observed pulse width evolution only becomes
apparent when examining durations that cover a broad temporal range.

{\bf Fig. 4.} - Haar wavelet scaleogram $\sigma_{X,\Delta t}$ versus
timescale $\Delta t$ (Panel A) and $\Delta t/t$ (Panel B) for the
ensemble of flares under study.  The expected level for Poisson noise
has been subtracted out.  Because $\sigma_{X,\Delta t}$ is calculated
from the natural logarithm of the XRT count rate, it can be interpreted
as a measure of RMS fractional flux variation versus timescale.
The scaleograms reach maximum and turn over on timescales $\Delta t
\approx 30-300$s and $\Delta t/t \approx 0.1-0.5$, indicating that the
flaring occurs on these characteristic timescales. Significant ($>3$-sigma
level) variability is observed on timescales $\Delta t\gtrsim 3$s and
$\Delta t/t \gtrsim 0.01$, however, $\sigma_{X,\Delta t} \propto \Delta t$
(dotted red curves) indicates that this variation is due to flaring on
intrinsically longer timescales.

{\bf Fig. 5.} - The flare rise time $T_{r}$ plotted vs. the time of peak
flux $T_{p}$.  As in Figure 3, multiple flares from individual GRBs are displayed with a unique color-symbol combination, whereas GRBs with only one flare are represented by a black diamond.  An increasing trend similar to that seen between $T_{90}$
and $T_{p}$ is evident in the data.  The observed rise times are
largely insensitive to the effects of background subtraction.


\bigskip

\section*{Figures}

\begin{figure}	\label{fig:all_flares}
\plotone{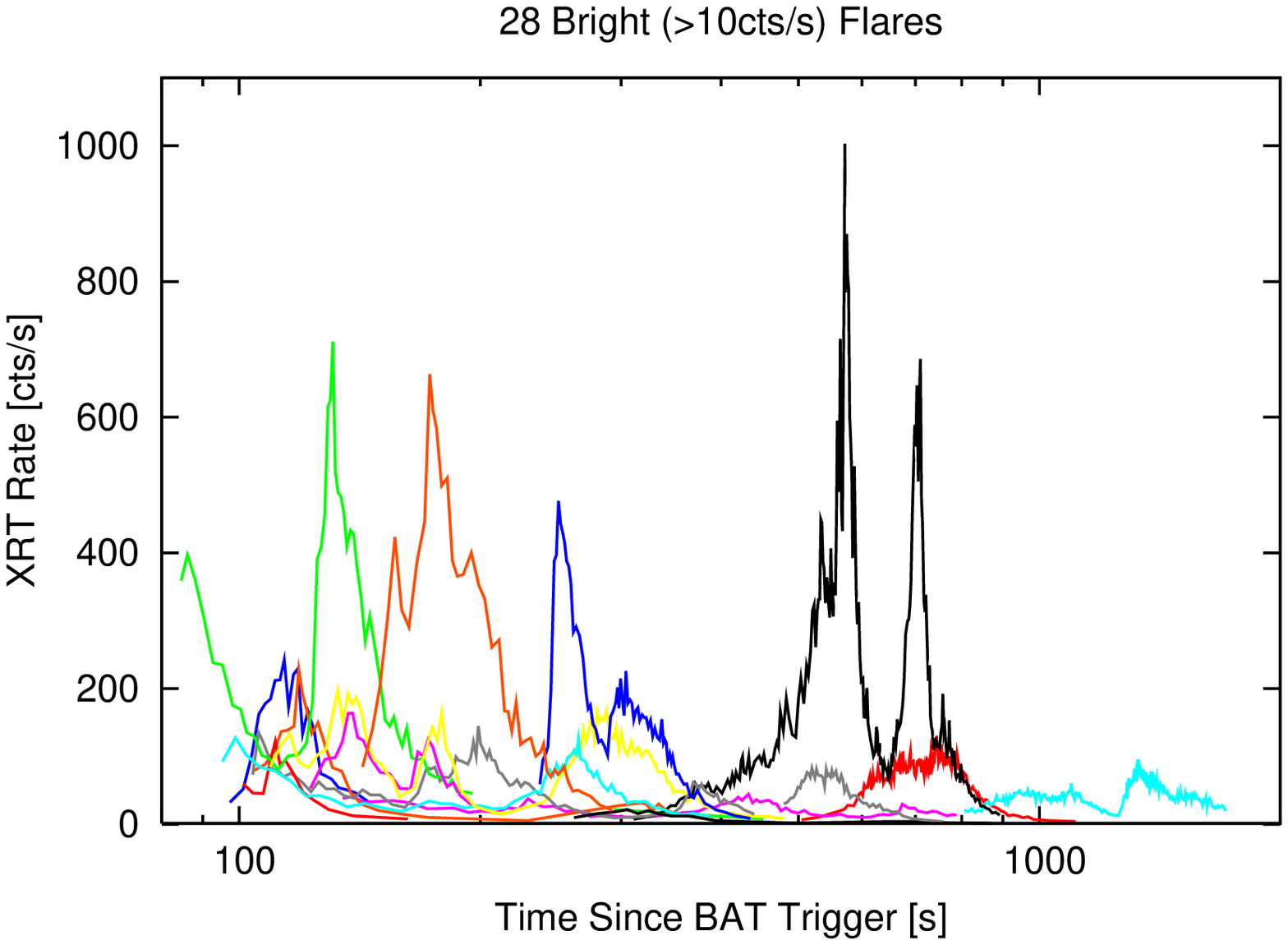}
\end{figure}

\clearpage

\begin{figure}	\label{fig:060714_dt}
\plotone{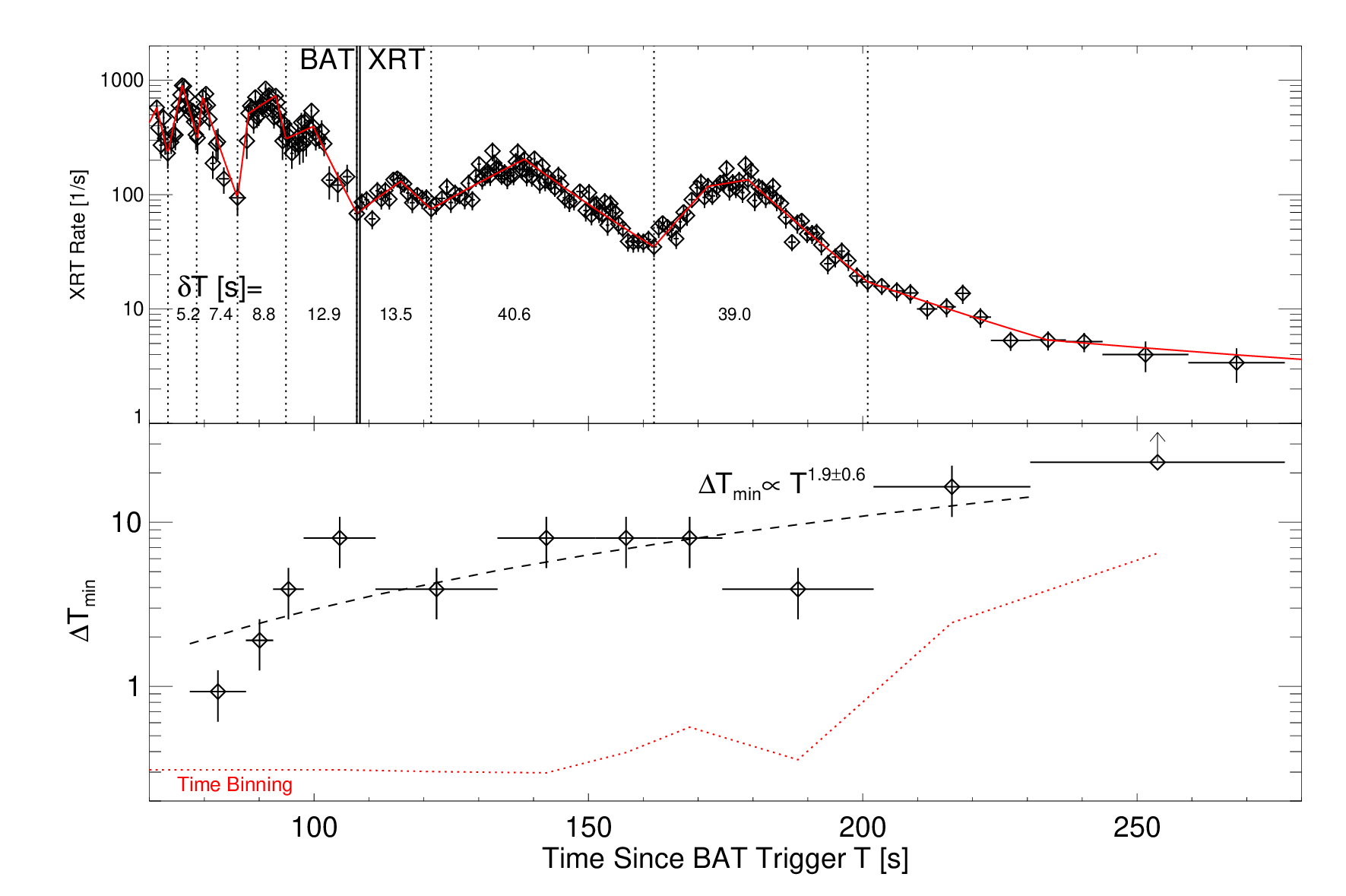}
\end{figure}

\clearpage

\begin{figure}	\label{fig:time_plot}
\plotone{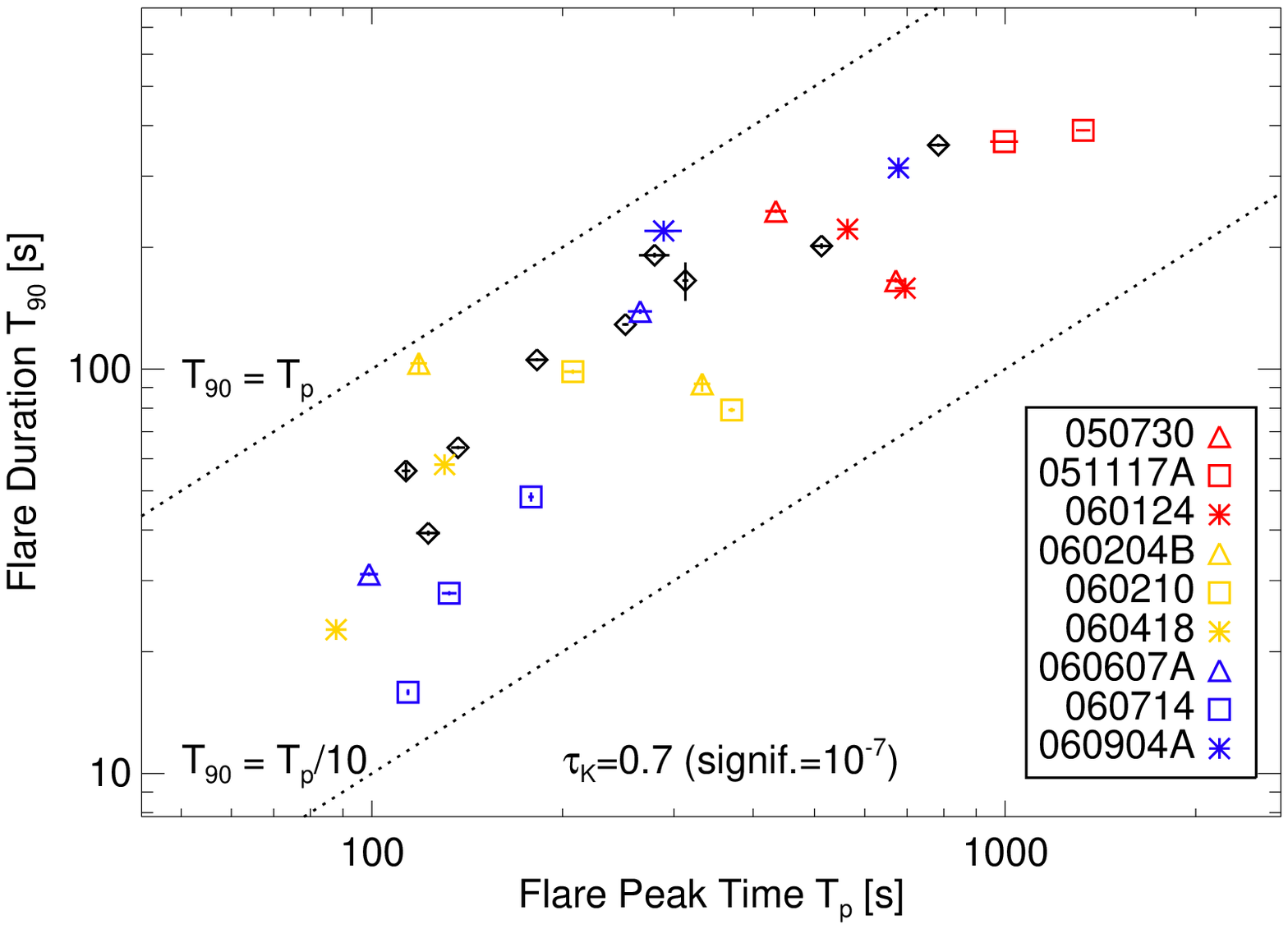}
\end{figure}

\clearpage


\begin{figure}[H] \label{fig:multi_haar}
\centerline{{\includegraphics[width=4.5in]{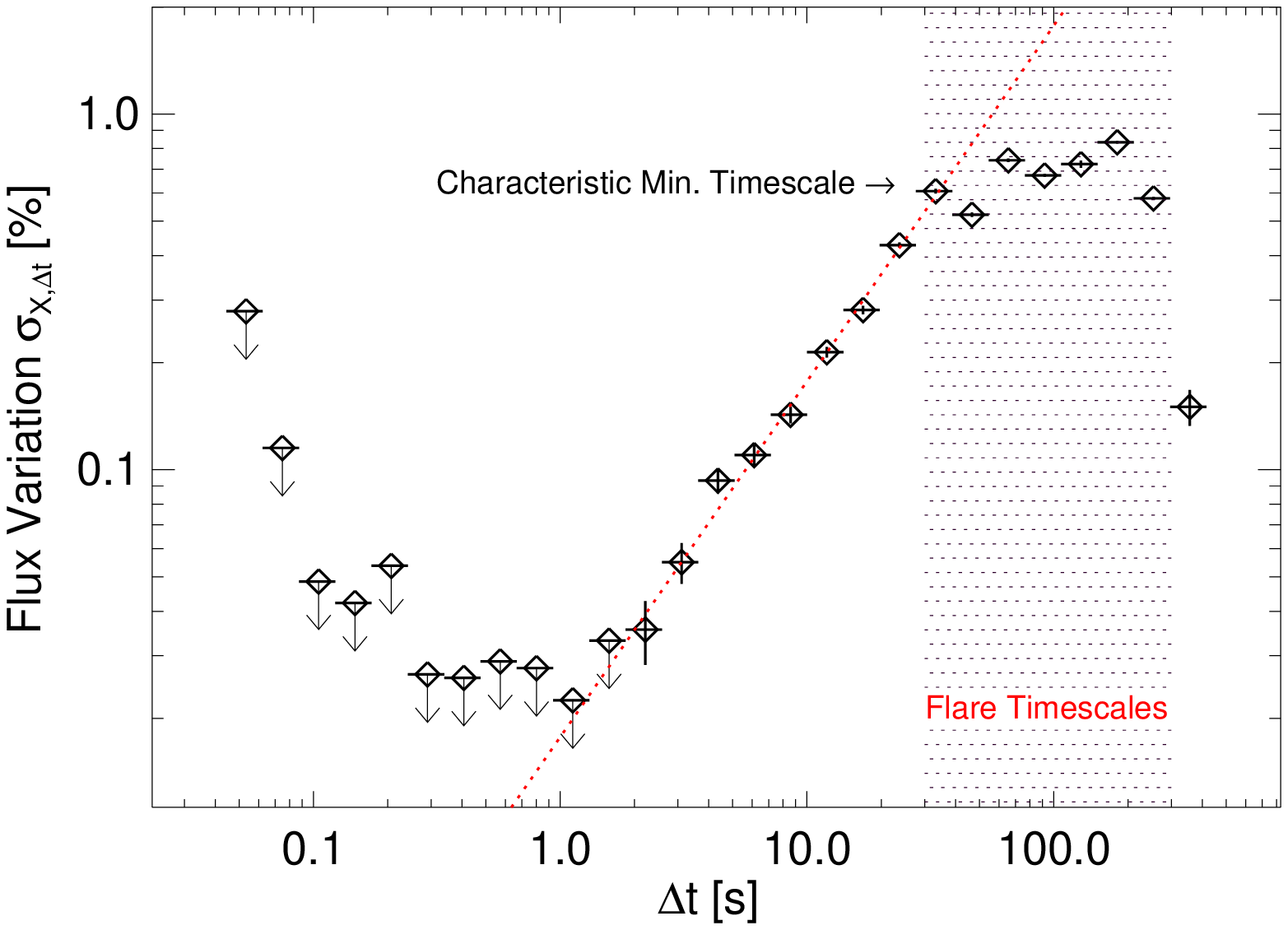}}}
\centerline{\includegraphics[width=4.5in]{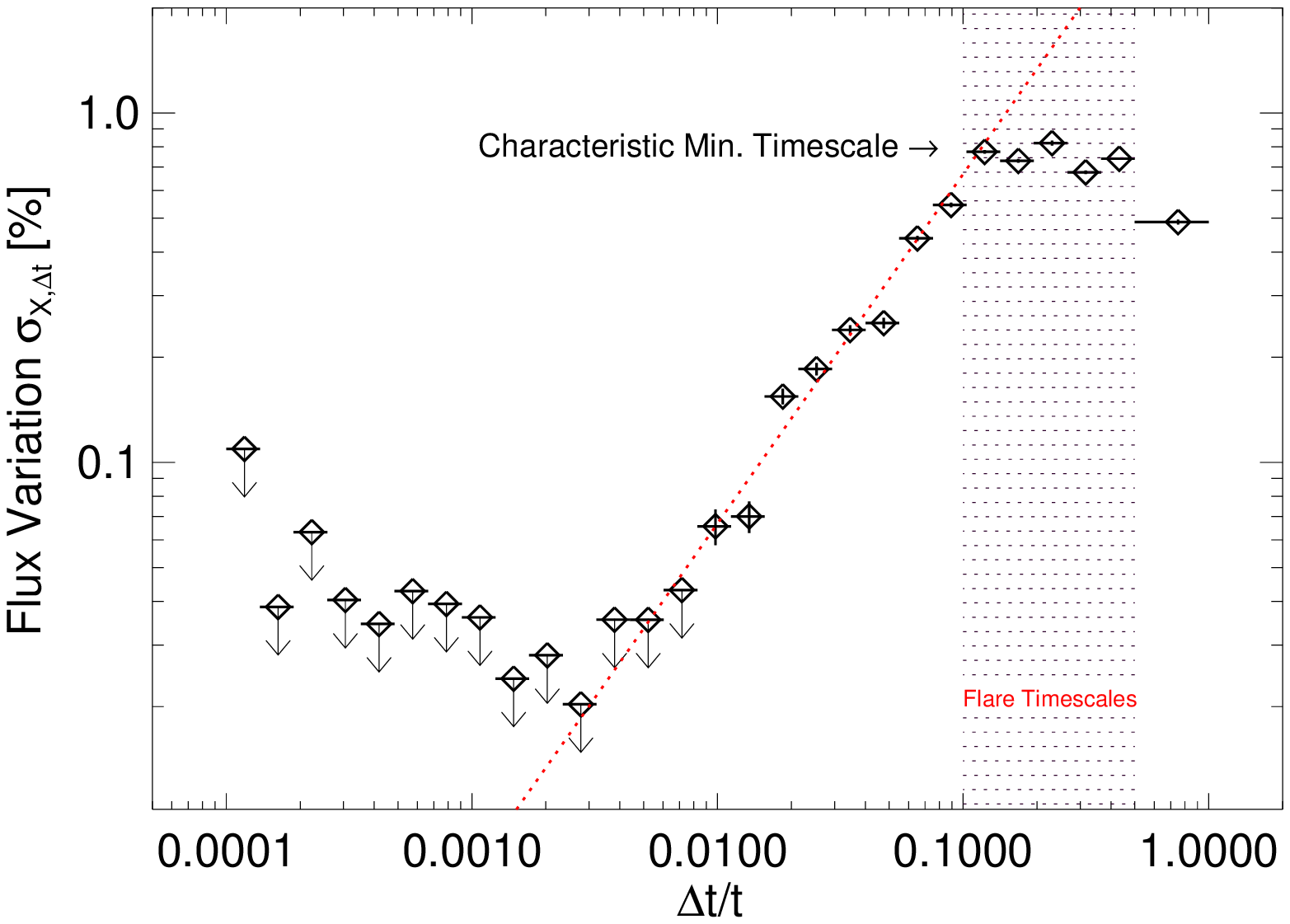}}
\end{figure}

\clearpage

\begin{figure}	\label{fig:time_plot_rise}
\plotone{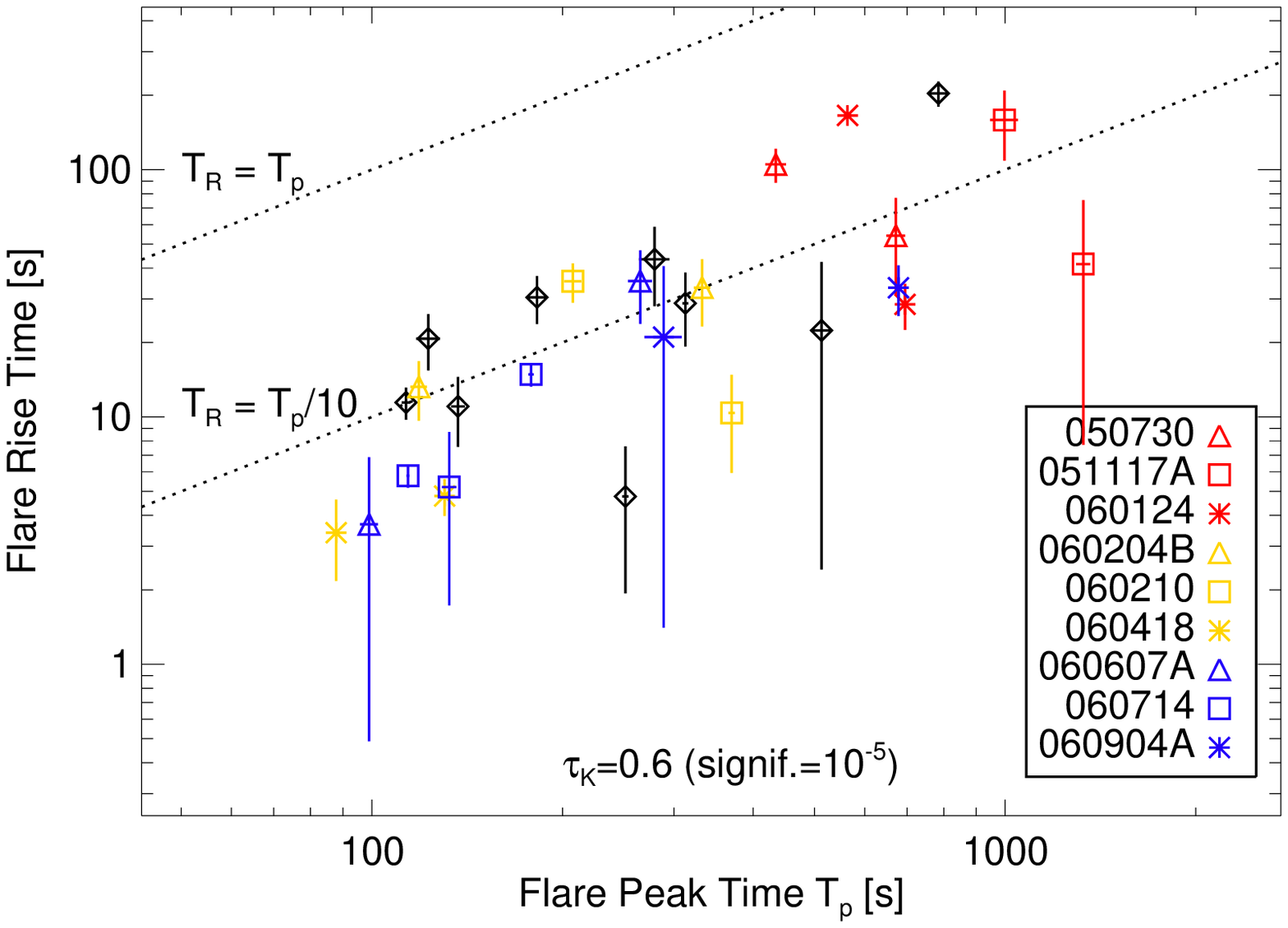}
\end{figure}

\clearpage


\begin{deluxetable}{rrrrrrrrr}

\label{Table:sample} \tablecolumns{5} \tablewidth{0pc}
\tablecaption{Pulse Duration \& Time of Peak Flux For Our Entire Sample}
\tablehead{ \colhead{GRB} & \colhead{Time Region} & \colhead{$T_{90}$}
& \colhead{$T_{\rm peak}$} & \colhead{$T_{\rm rise}$}
\\
\colhead{ } & \colhead{(s) } & \colhead{(s)}  & \colhead{(s) } &
\colhead{(s)}	}

\startdata

\\\hline
050502B & 400.0 -- 1200.0 & 358.1 $\pm$ 3.2 & 784.6 $\pm$ 23.7 & 203.5
$\pm$ 23.7 \\
050607 & 250.0 -- 600.0 & 165.5 $\pm$ 18.0 & 312.6 $\pm$ 3.2 & 28.8 $\pm$
9.6 \\
050713A & 95.0 -- 150.0 & 39.3 $\pm$ 0.5 & 122.8 $\pm$ 5.3 & 20.7 $\pm$
5.3 \\
060111A & 200.0 -- 500.0 & 191.3 $\pm$ 2.5 & 279.6 $\pm$ 15.4 & 43.4 $\pm$
15.4 \\
060312 & 100.0 -- 200.0 & 56.0 $\pm$ 2.5 & 113.2 $\pm$ 1.7 & 11.4 $\pm$
1.7 \\
060526 & 230.0 -- 450.0 & 128.9 $\pm$ 0.9 & 251.4 $\pm$ 2.8 & 4.8 $\pm$
2.8 \\
060604 & 120.0 -- 200.0 & 63.9 $\pm$ 0.5 & 136.8 $\pm$ 3.4 & 11.0 $\pm$
3.5 \\
060904B & 140.0 -- 300.0 & 105.4 $\pm$ 0.8 & 182.3 $\pm$ 6.6 & 30.5 $\pm$
6.7 \\
060929 & 470.0 -- 800.0 & 201.5 $\pm$ 3.0 & 513.0 $\pm$ 19.9 & 22.4 $\pm$
20.0 \\
050730 & 300.0 -- 600.0 & 245.6 $\pm$ 2.4 & 434.4 $\pm$ 16.3 & 105.1 $\pm$
16.5 \\
050730 & 600.0 -- 800.0 & 165.4 $\pm$ 1.7 & 672.2 $\pm$ 22.7 & 54.1 $\pm$
22.8 \\
051117A & 800.0 -- 1250.0 & 365.0 $\pm$ 1.7 & 997.1 $\pm$ 50.0 & 158.9
$\pm$ 50.1 \\
051117A & 1250.0 -- 1725.0 & 389.3 $\pm$ 1.4 & 1328.3 $\pm$ 33.8 &
41.5 $\pm$ 33.8 \\
060124 & 300.0 -- 650.0 & 221.7 $\pm$ 1.6 & 563.8 $\pm$ 7.9 & 165.7 $\pm$
7.7 \\
060124 & 650.0 -- 900.0 & 158.4 $\pm$ 1.4 & 694.9 $\pm$ 6.0 & 28.5 $\pm$
6.0 \\
060204B & 100.0 -- 270.0 & 103.3 $\pm$ 6.4 & 118.6 $\pm$ 3.5 & 13.2 $\pm$
3.6 \\
060204B & 270.0 -- 450.0 & 91.9 $\pm$ 3.9 & 332.3 $\pm$ 9.8 & 33.4 $\pm$
10.1 \\
060210 & 165.0 -- 300.0 & 98.5 $\pm$ 1.1 & 207.6 $\pm$ 6.5 & 35.4 $\pm$
6.4 \\
060210 & 350.0 -- 450.0 & 79.2 $\pm$ 0.9 & 369.9 $\pm$ 4.4 & 10.4 $\pm$
4.4 \\
060418 & 83.0 -- 110.0 & 22.7 $\pm$ 0.4 & 87.8 $\pm$ 1.2 & 3.4 $\pm$
1.2 \\
060418 & 122.0 -- 200.0 & 58.0 $\pm$ 0.6 & 130.3 $\pm$ 0.8 & 4.8 $\pm$
0.8 \\
060607A & 93.0 -- 130.0 & 31.1 $\pm$ 0.3 & 99.0 $\pm$ 3.2 & 3.7 $\pm$
3.2 \\
060607A & 220.0 -- 400.0 & 138.8 $\pm$ 1.9 & 265.3 $\pm$ 11.6 & 35.5 $\pm$
11.7 \\
060714 & 100.0 -- 125.0 & 15.9 $\pm$ 0.2 & 114.1 $\pm$ 0.6 & 5.8 $\pm$
0.6 \\
060714 & 125.0 -- 160.0 & 27.9 $\pm$ 0.3 & 132.5 $\pm$ 3.5 & 5.2 $\pm$
3.5 \\
060714 & 160.0 -- 230.0 & 48.3 $\pm$ 1.2 & 178.4 $\pm$ 1.7 & 14.9 $\pm$
1.6 \\
060904A & 250.0 -- 600.0 & 219.6 $\pm$ 12.8 & 288.9 $\pm$ 19.5 & 21.0
$\pm$ 19.6 \\
060904A & 600.0 -- 1000.0 & 314.2 $\pm$ 6.9 & 678.5 $\pm$ 7.7 & 33.3 $\pm$
7.7 \\ 
\enddata

\end{deluxetable}


\end{document}